\documentclass[conference]{IEEEtran}
\IEEEoverridecommandlockouts
\usepackage{cite}
\usepackage{amsmath,amssymb,amsfonts}
\usepackage{algorithmic}
\usepackage{graphicx}
\usepackage{textcomp}
\usepackage{xcolor}
\def\BibTeX{{\rm B\kern-.05em{\sc i\kern-.025em b}\kern-.08em
    T\kern-.1667em\lower.7ex\hbox{E}\kern-.125emX}}
\graphicspath{ {./Pic/} }
\begin{document}

\title{Implementing an Optimized and Secured Multimedia Streaming Protocol in a Participatory Sensing Scenario.}

\author{\IEEEauthorblockN{Vaiuso Andrea}
\IEEEauthorblockA{\textit{Researcher} \\
\textit{Zurich University of Applied Science}\\
Winterthur, Switzerland \\
vaiu@zhaw.ch}
}

\maketitle

\begin{abstract}
Multimedia streaming protocols are becoming increasingly popular in Crowdsensing due to their ability to deliver high-quality video content over the internet in real-time. Streaming multimedia content, as in the context of live video streaming, requires high bandwidth and large storage capacity to ensure a sufficient throughput. Crowdsensing can distribute information about shared video contents among multiple users in network, reducing storage capacity and computational and bandwidth requirements. However, Crowdsensing introduces several security constraints that must be taken into account to ensure the confidentiality, integrity, and availability of the data. In the specific case of video streaming, commonly named as visual crowdsensing (VCS) within this context, data is transmitted over wireless networks, making it vulnerable to security breaches and susceptible to eavesdropping and interception by attackers. Multimedias often contains sensitive user data and may be subject to various privacy laws, including data protection laws and laws related to photography and video recording, based on local GDPR (General Data Protection Regulation). For this reason the realization of a secure protocol optimized for a distributed data streaming in real-time becomes increasingly important in crowdsensing and smart-enviroment context. In this article, we will discuss the use of a symmetric AES-CTR encryption based protocol for securing data streaming over a crowd-sensed network.
\end{abstract}

\section{Crowdsourcing}

Crowdsourcing, sometimes referred to as participatory sourcing is a technique of data collecting that consists in the extraction of such data from a large number of individual users via mobile devices. These types of devices are often capable of sensing and computing, such as smartphones, tablets, smartwatches, and other wearable devices. In summary, crowdsourcing represents a versatile and distributed method that can be used to collect data for a wide range of tasks, including traffic management, environmental monitoring, health monitoring, disaster response, social media analysis, etc. \cite{guo2015mobile}. This kind of data extraction deeply depends on the credibility of the user involved to gather precise information. Indeed, there is no assurance that individuals will provide truthful or exact information about the environment. To tackle this problem, crowdsourcing is based on the use of incentives that can be given to the users in some ways, in order to motivate them to provide more accurate data. Also, implementing trust mechanisms is necessary to assess the credibility of the data given by each individual on the system. For this motivation, crowdsourcing applications can be grouped into four categories, and they are voting system, information sharing system, game and creative system \cite{6113213}.
\subsection{Smart environment through crowdsensing}
Mobile Crowd Sensing (MCS) is the most popular mobile crowdsourcing system. It uses mobile devices such as smartphones, wearable devices, and other portable devices to collect and produce high-level reliable output data as input for complex systems, such as in the \textit{smart environment} context. A smart environment is generally defined as an intelligent space (such as a city, university, farm, industry, etc.) that is equipped with sensors, actuators, and other devices that can collect and process data, and that can be controlled by an intelligent, centralized or decentralized distributed system that often uses artificial intelligence (AI) models to perform specific high-level tasks. Crowdsensing is commonly considered a powerful way to partially or completely replace the sensors phisically installed in the smart place with sensors present in users' devices. Crowdsensing is widely used in infrastructure applications due to its versatility, scalability, and low-cost implementation, and can perform tasks such as measuring large-scale events and conditions related to public infrastructure. For example, in smart cities this includes the measurement of traffic flow, road conditions, available parking spaces, and the identification of issues with public works such as malfunctioning fire hydrants or broken traffic lights. It also encompasses real-time tracking of transit systems to keep tabs on their locations and movements \cite{6069707}.  Indeed, the use of crowdsensing decrease the necessity of specialized sensor installation and utilization in favor of devices owned by users of the system. 

\subsection{Visual crowdsensing}
Visual crowdsensing (VCS) asks people to capture the details of interesting objects/views in the real world in the form of pictures or videos. It has recently attracted considerable attention due to the rich information that images and videos can provide thanks to AI based algorithms. Lot of previous project indicate that VCS is useful and in many cases superior to traditional visual sensing that relies on deployment of stationary cameras for monitoring \cite{guo2017emergence}. While VCS has shown great potential in the range of applications described above, there are several challenges associated using visual data contributed by crowdsensing. One of the most significant hurdles is the sheer volume of pictures and videos submitted by users, which can quickly become overwhelming. Furthermore, the quality and reliability of the data can vary widely, with some users contributing clear and accurate information, while others may not. To address these issues, researchers are exploring new approaches to manage and filter the large volume of data generated by VCS, implementing new techniques for earning incentives. 

Another challenge associated with VCS is the issue of redundant data, as multiple users may share similar pictures or videos. Put simply, these factors can create challenges when trying to analyze and understand the data. The visual data from the crowd contains extra details like geo-tags and context, which can be helpful but also add complexity to the analysis. Current approaches to Visual Crowd Sensing (VCS) heavily depend on the data itself to address these challenges. However, with the progress made in machine learning and data analytics, there are more opportunities to enhance the effectiveness and precision of VCS.

Finally, the exchange of video and images with VCS could involve large amounts of data being transmitted over the network, which could pose a security constraint in terms of privacy and message integrity, covered in section \ref{threats}.

\section{Threats in crowdsensing scenario}\label{threats}
Security in the context of Mobile crowdensing (MCS) refers to protecting the data and the systems from unauthorized access, use, modification, destruction, or other forms of interference. A secure MCS system, especially when visual data is involved, should be designed in a way that can withstand security breaches and prevent any unauthorized parties from gaining access to sensitive data or processing results. At the same time, the system should continue to function normally, ensuring that the data is protected at all times \cite{8080202}.

Crowdsensing and VSC involve the transmission of large amounts of data over the Internet. The use of collected data retrieved from users' personal devices implies the need to protect privacy by securing sensitive data during data transmission. Privacy concerns related to video streaming arise when personal information is captured or shared without consent, or when data are shared with third-party applications or service providers, particularly in situations where individuals may have a reasonable expectation of privacy, like in smart environment context. In addition, when privacy affects multimedia streaming informations, such as video or images shared over internet, it is difficult to ensure privacy due to the raw nature of these data. Ensuring user's privacy also involves \cite{yang2015security}:
\begin{itemize}
	\item \textbf{Data Integrity}: When data through crowdsensing are collected, there is a risk that someone with malicious intent may try to modify the data or add fake data to the system. To ensure the accuracy and reliability of the data, \textit{cryptographic techniques like digital signatures and message authentication codes are commonly used}. These methods can help to verify the authenticity of the data and ensure that it has not been modified either accidentally or voluntarily. 
	\item \textbf{User authentication}: It is sometimes necessary to authenticate crowd-sourcing participants before they join the network to assure the authenticity of the data. However, due to the large number of crowd sourcing participants, it is not efficient to let end users perform authentication. Therefore, \textit{a distributed authentication mechanism is desired} to provide efficient and reliable authentication service for crowdsourcing participants.
	\item \textbf{Channel security}: Crowdsensing data is transmitted over wireless networks, which are susceptible to eavesdropping and interception by attackers. To ensure the security of the communication channels, \textit{encryption techniques must be used} to protect the confidentiality of the data during transmission.
\end{itemize}

\section{Featured works and state of the art}
Data collection must implement a secure mechanism for data transfer from user to a service provider (SP). This process includes various security measures such as encryption, provenance authentication, secure routing, and key exchange. When data is conveyed through a mobile crowd sensing (MCS) system, additional protection measures such as encryption, data cloak, and data generalization are employed to ensure confidentiality, integrity, and provenance authentication. In a centralized MCS architecture, data is typically transmitted directly to the SP, and current efforts are focused on utilizing data encryption to enhance security. 

In \cite{wu2013k}, data generalization is applied to support \textit{k-anonymity}. In this scheme, users change their pseudonym periodically. The worker generates a new key pair for this pseudonym, and a trusted authority called reputation and pseudonym manager (RPM) is introduced to sign the public key by applying a blind RSA signature mechanism to provide Au for the pseudonym and key pair. The main problem in this approach, when multimedia data are involved, is that no protection are provided to users and other people who appear in shared videos/pictures. 

In Qiu et al \cite{6883146}, they introduced a privacy-preserving approach called SLICER for crowdsourcing of multimedia data. SLICER is among the first k-anonymous schemes that offer robust privacy protection for participants while ensuring data quality. To achieve this, SLICER uses a data coding technique and message transfer strategies. However, SLICER's use of RSA asymmetric encryption is not practical for encrypting large amounts of data, and its evaluation does not consider the performance of high-quality multimedia streaming.

Our paper aims to introduce a framework based on secured RTP protocol that is specifically designed for privacy-preserving multimedia streaming. This framework addresses the limitations of related methods by offering more efficient symmetric encryption method that can handle large data volumes and considering the performance of high-quality multimedia streaming. Indeed, RTP protocol is a commonly used method for transporting multimedia streaming data, and this make our framework easely compatible with many well-known multimedia software streaming applications.

\subsection{RTSP and RTP}
RTSP (Real Time Streaming Protocol) [RFC 7826], RTP (Real-time Transport Protocol) [RFC 3550], and SRTP (Secure Real-time Transport Protocol) [RFC 3711] are three protocols developed by IETF (Internet Engineering Task Force) and documented in RFC (Request for comment), that are commonly used in streaming real-time applications. Specifically, RTSP is an application-layer protocol for the setup and delivery control of data with real-time properties. RTSP provides an extensible framework to enable controlled, on-demand delivery of real-time data, such as audio and video. RTSP is used to negotiate an RTP socket, used for actually transmitting the content over the network: RTP provides an end-to-end network transport functions, based on UDP transport protocol, suitable for applications transmitting real-time data over multicast or unicast network services. RTP does not address resource reservation and does not guarantee quality-of-service for real-time services, but is designed to deliver the best data throughput and provide encoding/decoding, encryption and multimedia setup functionality. RTP is commonly used in conjunction with RTCP (RTP Control Protocol), which monitors the quality of service and carries information regarding the participants in a session.
\subsection{SRTP}
SRTP (Secure Real-time Transport Protocol) was developed with the aim of improving the security objectives inherent in RTP (Real-time Transport Protocol) while preserving high throughput constraints. In contrast to RTP, which provides no security features and is vulnerable to various attacks such as eavesdropping and tampering, SRTP provides security features such as symmetric AES-CTR encryption, message integrity and authentication with HMAC. These protocols are commonly used in a variety of streaming applications, including video conferencing, multimedia streaming, and any other real-time data transfer that requires secure transmission. With SRTP, these applications can achieve both high throughput and strong security, which is essential in modern communication systems where privacy and data protection are critical concerns.

\section{A python library for multimedia secured streaming in crowdsensing scenario}
The main purpose of this article is to introduce and provide a secured open source SRTP-based protocol that has been specifically designed for multimedia crowdsensing. Crowdsensing is a typical scenario where sensitive data are shared between untrusted parties, such as users themselves, and third-party service providers. In addition, the negotiation of TLS or IPSec session could result infeasible between untrusted peers. In this scenario a secure, computationally lightweight and real-time based protocol for fast data exchange is needed. Our work includes the realization of an open source protocol based on SRTP symmetric encryption/decryption AES-CTR (Advanced Encryption Standard \cite{miller2009advanced} in Counter operating Mode \cite{lipmaa2000ctr}) schema and HMAC message authentication code based on secure hash aglorithm SHA \cite{eastlake2001us}. This provides an extensible framework for encrypted multimedia exchange between untrusted parties upstream of a symmetric keys distribution operated by a third trusted party. 
\subsection{Testing framework}
The source code provided is written in Python 3.9 and utilizes several open-source Python libraries for connection management and efficient encryption/decryption and image manipulation. These libraries include PyCryptodome, HMAC and OpenCV. To demonstrate the efficiency of the code in practical use cases, extensive testing was carried out on a 2015 Macbook Pro machine. The tests involved analyzing the performance of the code while streaming two different types of MP4 local videos, which were encoded in both raw and JPEG formats, and varied in multiple resolutions (480p, 720p, and 1080p). The tests were designed to ensure that the code could handle a wide range of video formats and resolutions, while still maintaining a high level of performance and efficiency.
\subsection{Architecture overview}
In this section we will describe the system architecture: entities involved, protocols stack, message exchanged, key exchange mechanism, session key generation, encoding/decoding mechanism, encryption/decryption mechanism and integrity check mechanism:
\begin{itemize}
	\item \textbf{Entities involved}: We consider the following to be involved in the multimedia exchange: workers (system users, service providers, etc.) and a trusted third party to handle the key exchange. We will consider a scenario where a Service Provider (SP) requests a multimedia stream identified by a specific URI from a user in the crowd (A), but this kind of multimedia exchange can also work with user to user (A to B), or between multiple SPs ($SP_1$ to $SP_2$).
	\item \textbf{Protocols}: Our approach consists in two main protocols based in RTSP and S/RTP application protocols. SP requester can send a RTSP request message to A in order to ask for a multimedia stream, identified by ID, through SETUP method. This method must contains the following parameters:
	\begin{itemize}
		\item \textbf{URI}: indicates resource ID. This id can be fixed on a crowdsensing scenario, depending on type of live stream data requested.
		\item \textbf{Transport}: indicates transport mode. The developed method for this implementation is RTP - unicast. RTP port is needed to open RTP connection. RTP port can be fixed in a crowdsensing scenario with default multimedia port configurations.
		\item \textbf{Master-Salt}: a random, single-use value used as shared master salt.
	\end{itemize}
	Response SETUP request must contains a session ID, used for encryption. \\
	From this point, if a connection was successfully established, an SP can request the media stream by sending PLAY request. When PLAY request is received, the user can send by RTP configured socket the secured RTP stream to the SP. Other methods, namely PAUSE and TEARDOWN, are also available for respectively pausing and disconnecting the stream.
	\item \textbf{Master Session Key Exchange}: To ensure the secure transmission of SETUP requests, it's necessary to exchange a secret key and certificate in order to establish a secure TLS connection for SRTP requests and responses. Additionally, for RTP payload encryption and authentication, a secret shared key is required. In this scenario, a fixed Diffie-Hellman with an x509 public RSA-PSS certificate can be utilized, but the choiche of the best key exchange mechanism must be referred to the system architecture and actors. For this motivation, key exchange mechanism is not included in the open source code. In any case, it's important to assign the responsibility of providing the key and certificate to a trusted third party, such as a certification authority or a trusted service provider.
	\item \textbf{Key derivation}: Following S/RTP guidelines, session keys are derived from master session key using master salt and session ID:
	\[
IV_{PRF}=master\_salt\;\oplus\;session\_id\;\oplus\;2^{16}
	\]
	\[
	ks = PRF(IV_{PRF})
	\]
	Where $ks$ is a keystream where keys are derived, while $PRF$, is a pseudo-random function obtained from the execution of AES-CTR algorithm using $IV_{PRF}$ as first counter value. 3 keys are obtained from $ks$ first 768 bit, which are 256 bit session key ($Se_k$), 256 bit, salting key ($Sa_k$) and 256 bit authentication key ($Au_k$).
	\item \textbf{Frame encoding/decoding}: Frame encoding and decoding is obtained via OpenCV frame encoding function (\texttt{imencode}), using JPEG format. Finally, encoded data are formatted in base64.
	\item \textbf{Payload encryption/decryption}: Encoded frame is encrypted using AES-CTR using $Se_k$. Initial value for counter is obtained as:
	\[
	IV_{AES}=Sa_k\;\oplus\;session\_id\;\oplus\;2^{16}
	\]
	\[
	ks' = AES(IV_{AES})
	\]
	Finally, $ks'$ first $n$ bits are used for xoring $n$ bit encoded data frame:
	\[
	C=MSB(n,ks')\;\oplus\;data
	\]
	\item \textbf{Integrity check}: Finally, HMAC-SHA-1-256 is used to generate TAG on ciphered message, using $Au_k$ as MAC key. TAG value is tailed to RTP packet. The final packet is sent to the requester, where TAG value are verified using symmetric $Au_k$ key.
\end{itemize}

\subsection{Results}
Now we will have a look at the results we had with different video, encoding and key size.
\begin{figure}[]
  \includegraphics[width=\linewidth]{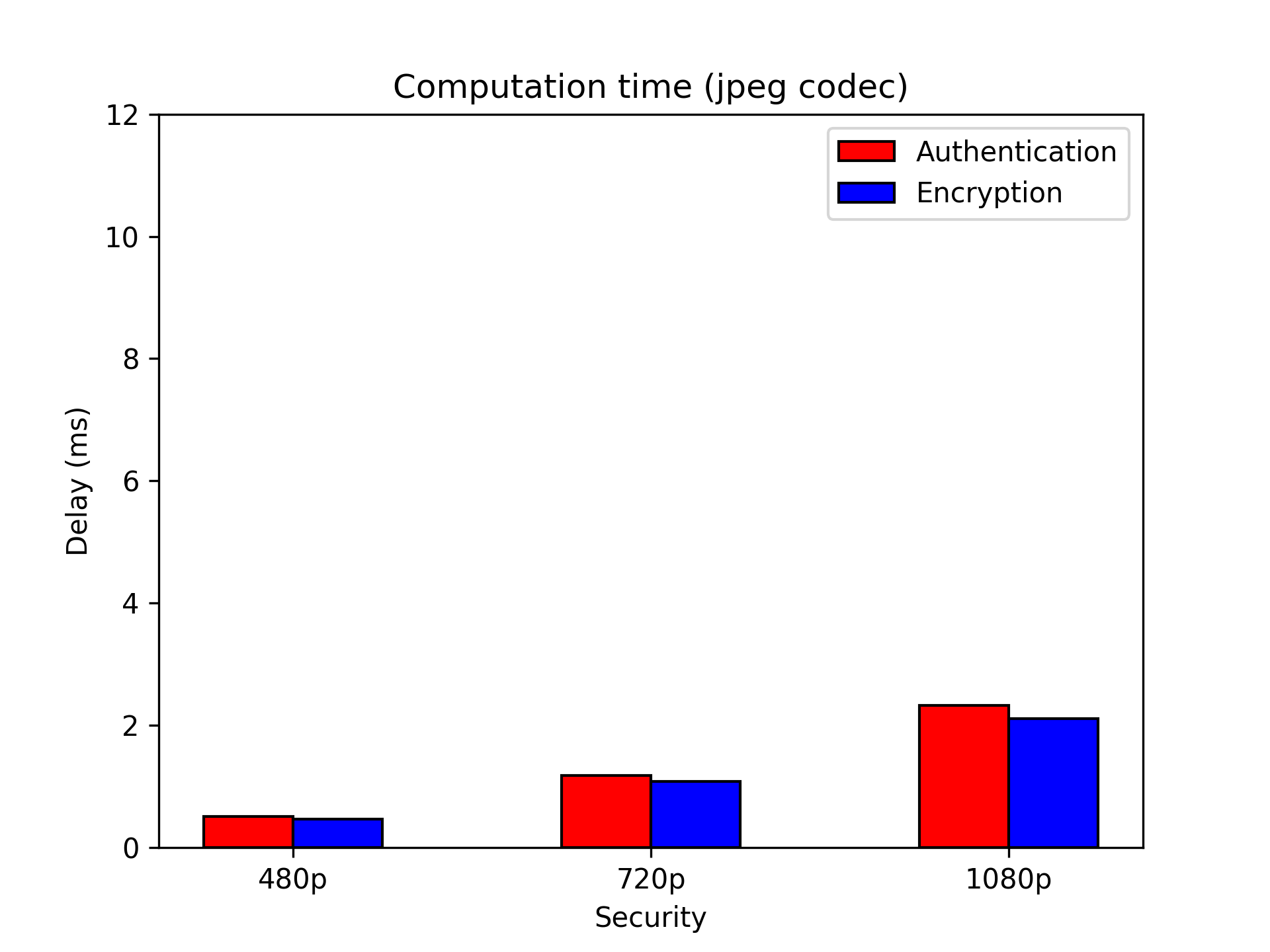}
  \caption{Computation time on client with multimedia transfer using JPEG coding, varying resolution.}\label{f1}
\end{figure}
\begin{figure}[]
  \includegraphics[width=\linewidth]{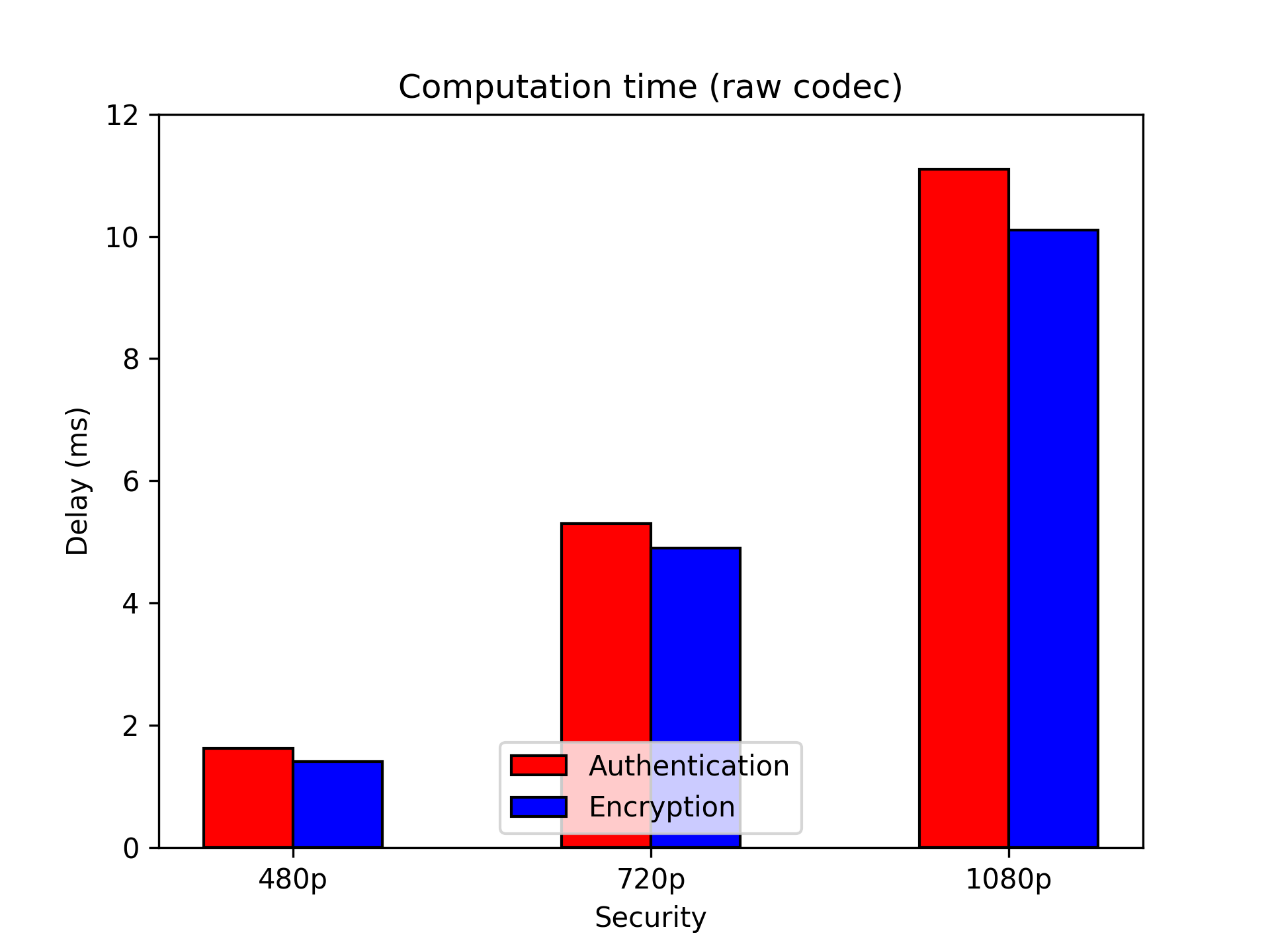}
  \caption{Computation time on client with multimedia transfer using no encoding, varying resolution.}\label{f2}
\end{figure}
The graphs presented in Figure \ref{f1} and \ref{f2} provide information on the variable nature of computation time when different frame sizes and encoding algorithms are used. The data presented in this analysis were obtained by calculating the average of all execution times for both encryption and authentication processes. Figures suggest that the choice of encoding algorithm can have a significant impact on the delay time when encryption and authentication is performed. It is noteworthy that using JPEG encoding results in significantly lower delay times: when using RAW encoding, the delay time varied from about 2 to 10 milliseconds, whereas JPEG encoding resulted in a delay time ranging from 0.5 to 2.5 milliseconds. These results suggest that the choice of encoding algorithm is a crucial factor in determining the efficiency of encrypting and authenticating video frames.
\begin{figure}[]
  \includegraphics[width=\linewidth]{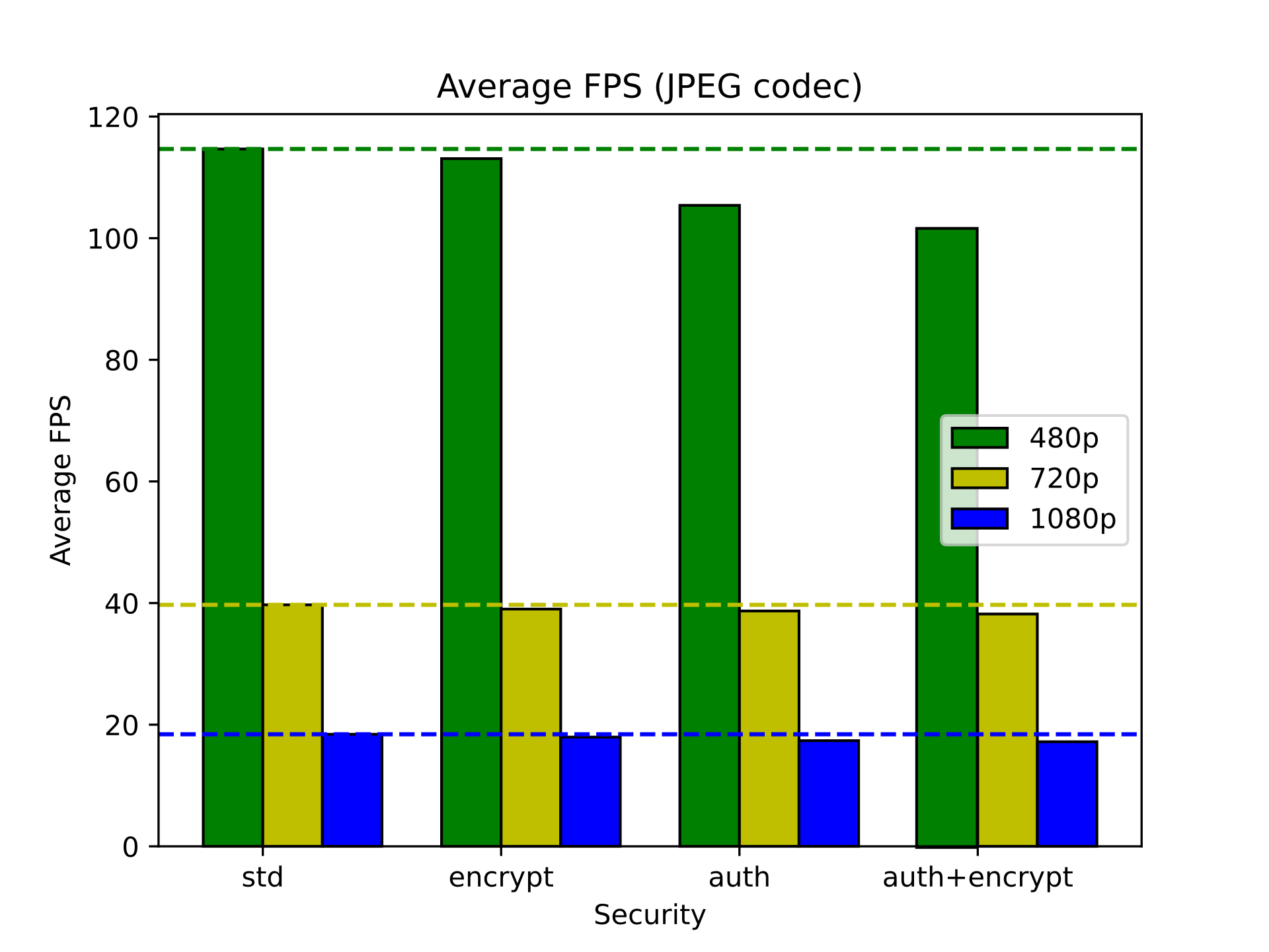}
  \caption{FPS performance on client with multimedia transfer using JPEG coding, varying resolution.}\label{f3}
\end{figure}
\begin{figure}[]
  \includegraphics[width=\linewidth]{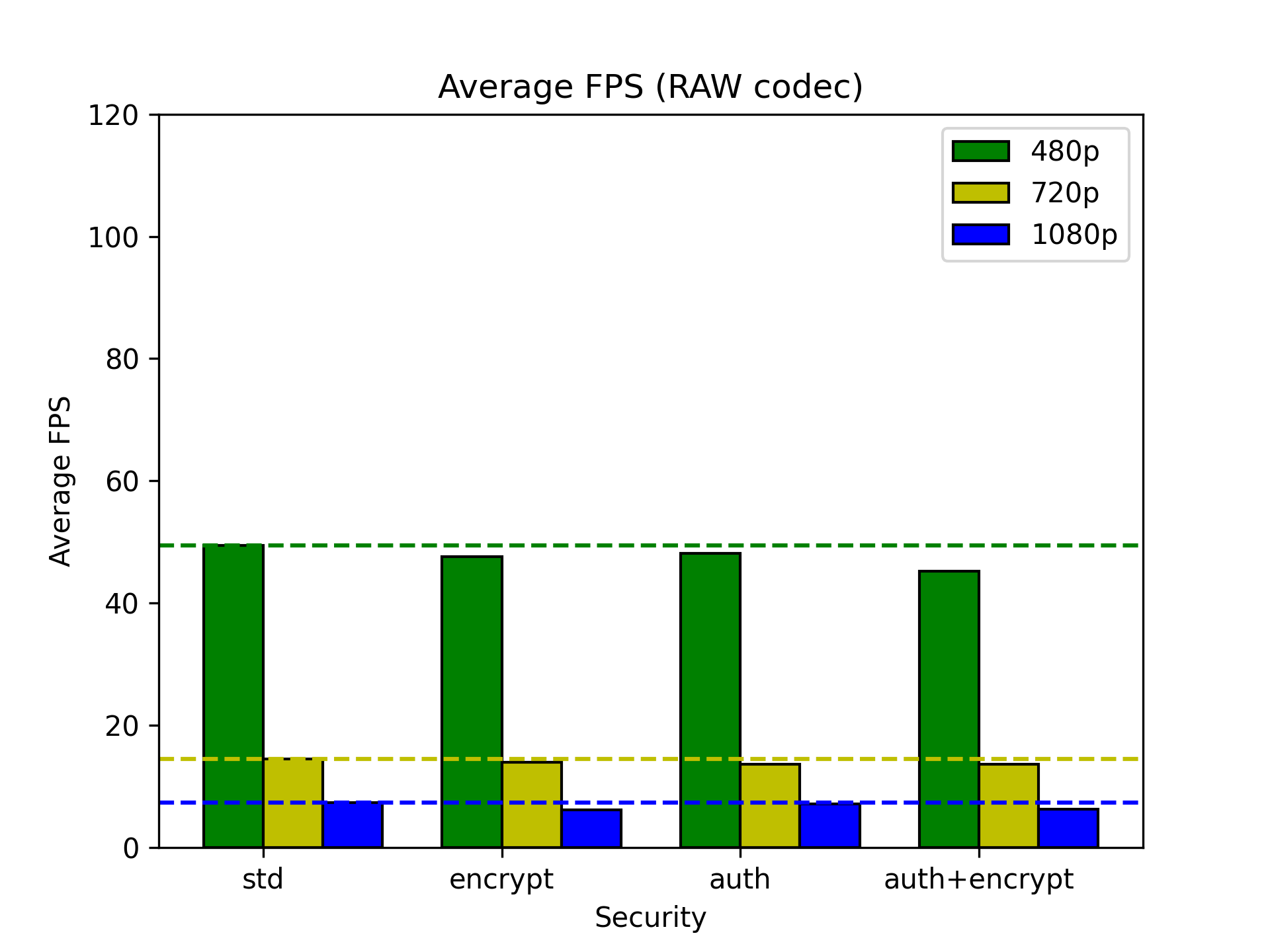}
  \caption{FPS performance on client with multimedia transfer using no encoding, varying resolution.}\label{f4}
\end{figure}

In addition to evaluating computation time, we also analyzed the average Frames Per Second (FPS) performance of the system while varying the encoding, image size, and security functionalities. The results presented in Figure \ref{f3} and \ref{f4} provide valuable insights on FPS performance while varying video encoding format. Once again, we observed that using JPEG encoding led to better FPS performance compared to not using any encoding. However, the most significant finding was the FPS differences between unsecured mode and authenticated encryption mode. We found that in both JPEG and RAW video encoding, the introduction of encryption and authentication only led to a difference of approximately \texttildelow 2 FPS. This suggests that there is no compelling reason to use unsecured stream transfer, as encryption and authentication introduce only a minimal overhead.

\begin{figure}[]
  \includegraphics[width=\linewidth]{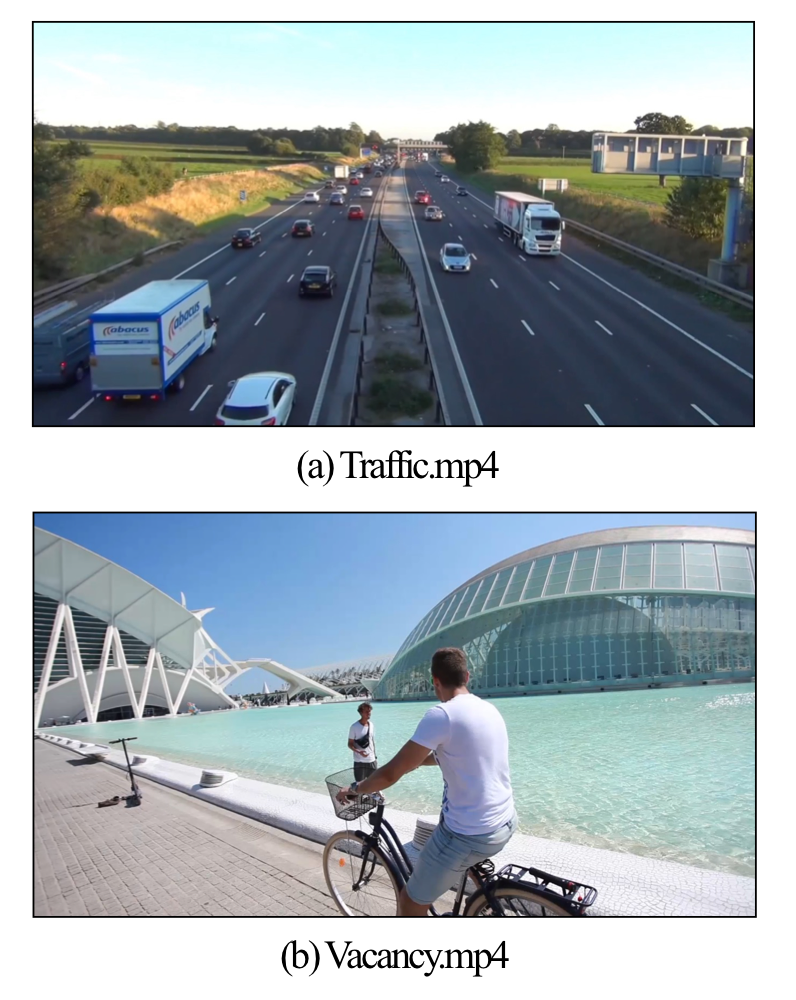}
  \caption{Video source frame example used for testing: (a) a static video capture by a CCTV system and (b) an HD video taken from a smartphone.}\label{f5}
\end{figure}

\begin{figure}[]
  \includegraphics[width=\linewidth]{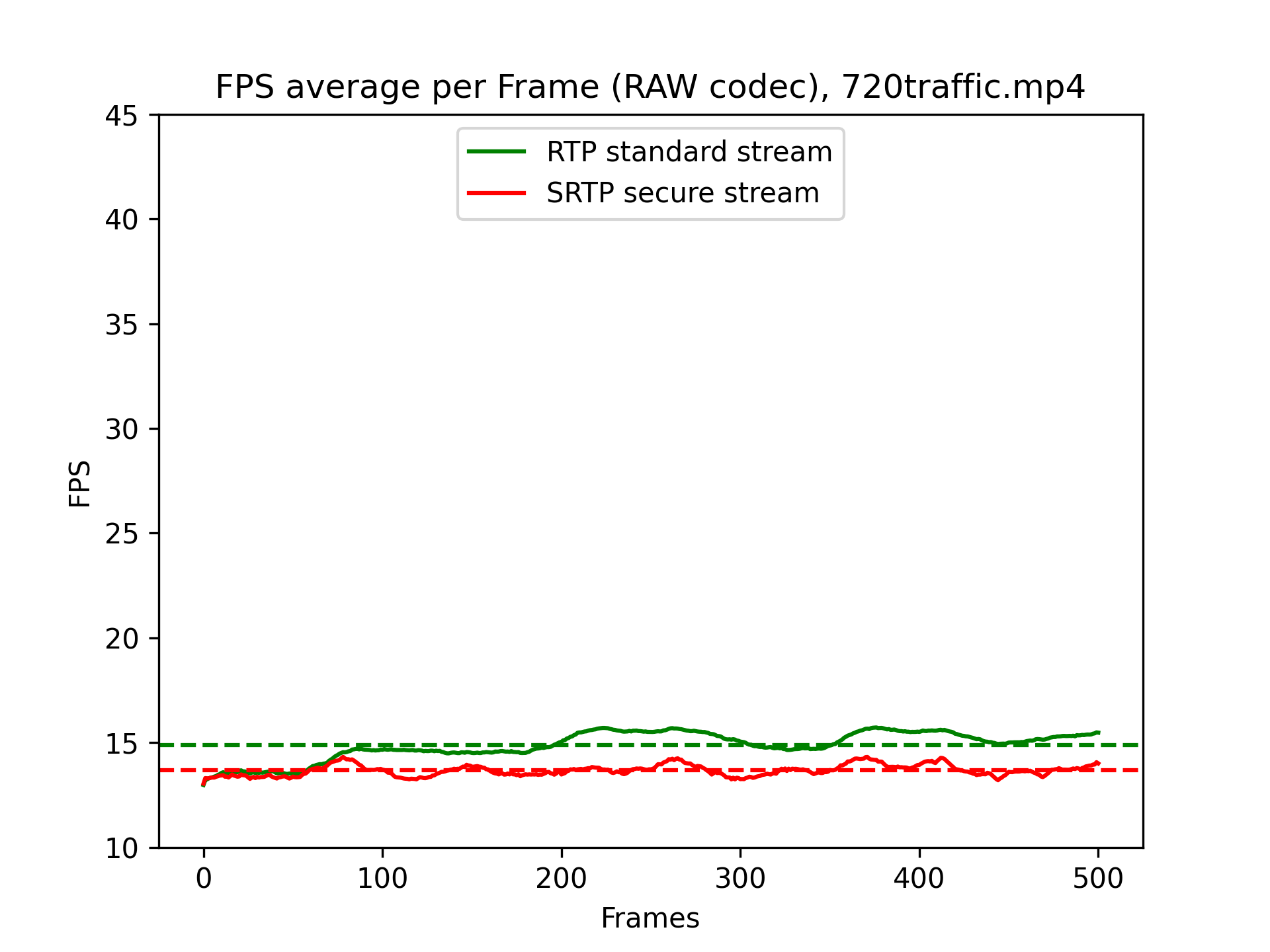}
  \caption{FPS performance averaged on video (a).}\label{f6}
\end{figure}

\begin{figure}[]
  \includegraphics[width=\linewidth]{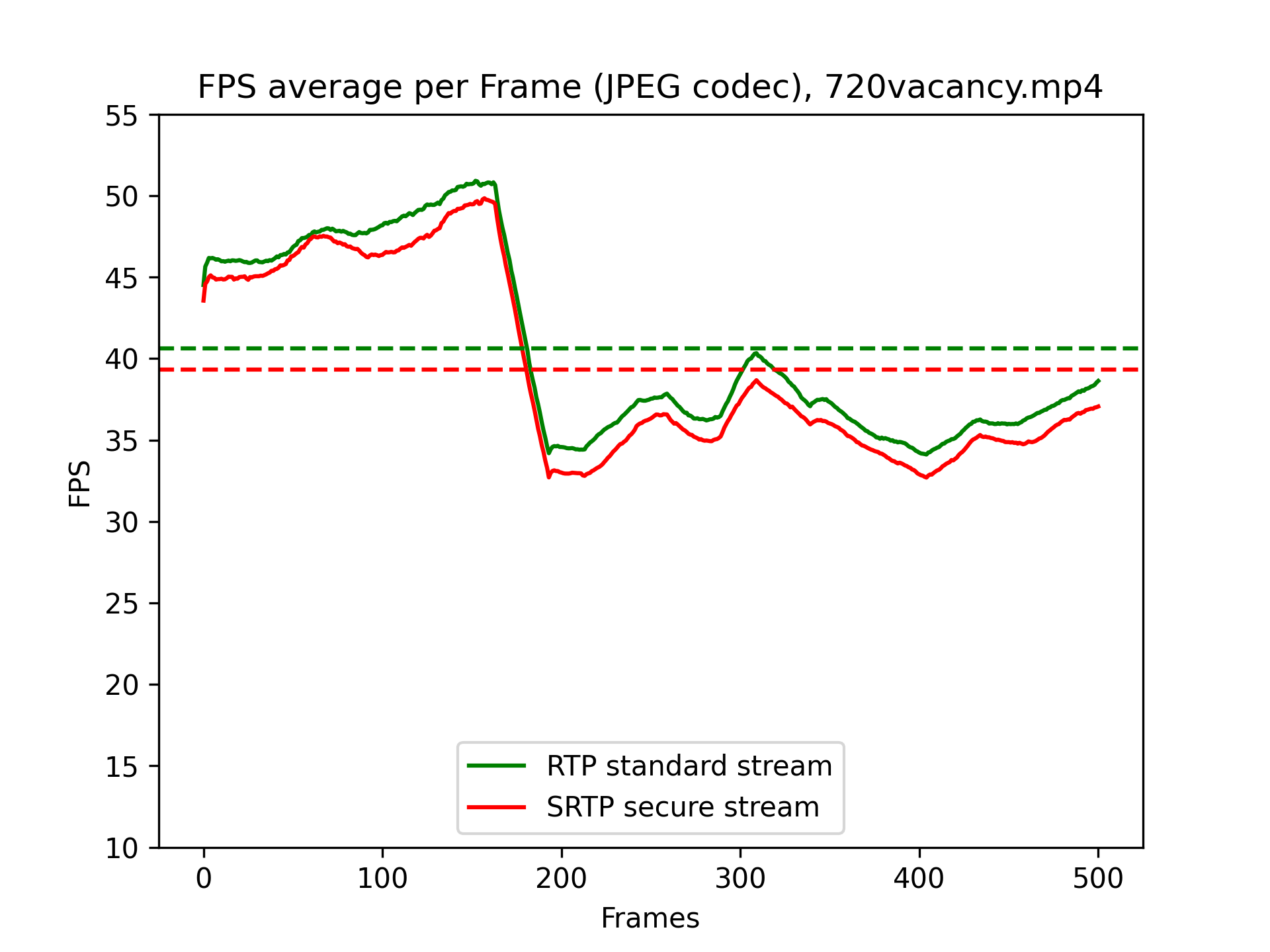}
  \caption{FPS performance averaged on video (b).}\label{f7}
\end{figure}

We also assessed the overall performance of the algorithm using video sources of different natures. Specifically, we used a static framing-based video of traffic flow, captured from a CCTV system, and a video example taken from a smartphone. By evaluating the average FPS variation during video playback, as shown in Figures \ref{f6} and \ref{f7}, we were able to compare the system's performance under different video sources.

Remarkably, our findings indicate that the FPS difference between unsecured and secured stream is just around \texttildelow 2 FPS, irrespective of the video source and codec. While the static video (Figure \ref{f6}) shows a smoother trend due to the relatively stable image information, the video captured with a smartphone (Figure \ref{f7}) shows a more unstable trend. This instability on client is caused by the non-implementation of the control protocol RTCP, which is necessary for the sending rate management. However, an RTCP implementation is typically required only when playing static videos at the right frame rate is needed. Within the context of our study, videos were used as input to an AI algorithm, or anyway received in live streaming, making RTCP implementation unnecessary.


\section{Conclusions and future works}
These results demonstrate the robustness and versatility of our S/RTP based algorithm in securing videos from different sources with comparable efficiency and effectiveness. Furthermore, these findings have practical implications for organizations in various industries that require secure video processing in crowdsensing scenario while ensuring privacy and performance.

\bibliographystyle{plain}
\bibliography{biblio.bib}

\begin{thebibliography}{10}

\bibitem{eastlake2001us}
D~Eastlake~3rd and Paul Jones.
\newblock Us secure hash algorithm 1 (sha1).
\newblock Technical report, 2001.

\bibitem{8080202}
Wei Feng, Zheng Yan, Hengrun Zhang, Kai Zeng, Yu~Xiao, and Y.~Thomas Hou.
\newblock A survey on security, privacy, and trust in mobile crowdsourcing.
\newblock {\em IEEE Internet of Things Journal}, 5(4):2971--2992, 2018.

\bibitem{6069707}
Raghu~K. Ganti, Fan Ye, and Hui Lei.
\newblock Mobile crowdsensing: current state and future challenges.
\newblock {\em IEEE Communications Magazine}, 49(11):32--39, 2011.

\bibitem{guo2017emergence}
Bin Guo, Qi~Han, Huihui Chen, Longfei Shangguan, Zimu Zhou, and Zhiwen Yu.
\newblock The emergence of visual crowdsensing: Challenges and opportunities.
\newblock {\em IEEE Communications Surveys \& Tutorials}, 19(4):2526--2543,
  2017.

\bibitem{guo2015mobile}
Bin Guo, Zhu Wang, Zhiwen Yu, Yu~Wang, Neil~Y Yen, Runhe Huang, and Xingshe
  Zhou.
\newblock Mobile crowd sensing and computing: The review of an emerging
  human-powered sensing paradigm.
\newblock {\em ACM computing surveys (CSUR)}, 48(1):1--31, 2015.

\bibitem{lipmaa2000ctr}
Helger Lipmaa, Phillip Rogaway, and David Wagner.
\newblock Ctr-mode encryption.
\newblock In {\em First NIST Workshop on Modes of Operation}, volume~39.
  Citeseer. MD, 2000.

\bibitem{miller2009advanced}
Frederic~P Miller, Agnes~F Vandome, and John McBrewster.
\newblock {\em Advanced encryption standard}.
\newblock Alpha Press, 2009.

\bibitem{6883146}
Fudong Qiu, Fan Wu, and Guihai Chen.
\newblock Privacy and quality preserving multimedia data aggregation for
  participatory sensing systems.
\newblock {\em IEEE Transactions on Mobile Computing}, 14(6):1287--1300, 2015.

\bibitem{wu2013k}
Sai Wu, Xiaoli Wang, Sheng Wang, Zhenjie Zhang, and Anthony~KH Tung.
\newblock K-anonymity for crowdsourcing database.
\newblock {\em IEEE Transactions on Knowledge and Data Engineering},
  26(9):2207--2221, 2013.

\bibitem{yang2015security}
Kan Yang, Kuan Zhang, Ju~Ren, and Xuemin Shen.
\newblock Security and privacy in mobile crowdsourcing networks: challenges and
  opportunities.
\newblock {\em IEEE communications magazine}, 53(8):75--81, 2015.

\bibitem{6113213}
Man-Ching Yuen, Irwin King, and Kwong-Sak Leung.
\newblock A survey of crowdsourcing systems.
\newblock In {\em 2011 IEEE Third International Conference on Privacy,
  Security, Risk and Trust and 2011 IEEE Third International Conference on
  Social Computing}, pages 766--773, 2011.

\end{thebibliography}

\end{document}